# Propagation Distance Required to Reach Steady-State Detonation Velocity in Finite-Sized Charges

Jianling Li[*], XiaoCheng Mi[#] and Andrew J. Higgins[#]

[*]National Key Laboratory of Shock Wave and Detonation Physics,
Institute of Fluid Physics, China Academy of Engineering Physics, Mianyang, Sichuan, China

[#]McGill University, Montreal, Quebec, Canada

**Abstract.** The decay of a detonation wave from its initial CJ velocity to its final, steady state velocity upon encountering a finite thickness or diameter charge is investigated numerically and theoretically. The numerical simulations use an ideal gas equation of state and pressure dependent reaction rate in order to ensure a stable wave structure. The confinement is also treated as an ideal gas with variable impedance. The velocity decay along the centerline is extracted from the simulations and compared to predictions base on a front evolution equation that uses the steady state detonation velocity-front curvature relation ($D_n$-κ). This model fails to capture the finite signaling speed of the leading rarefaction resulting from the interaction with the yielding confinement. This signaling speed is verified to be the maximum signal velocity occurring in the ideal ZND wave structure of the initial CJ velocity. A simple heuristic model based on the rarefaction generated by a one-dimensional interaction between the post-shock state and the confinement is proposed to provide an approximate description of the rest of the relaxation to steady state velocity.

## Introduction

A detonation in a sufficiently large charge will propagate at a velocity at or very near the ideal Chapman Jouguet (CJ) detonation velocity of the explosive. Upon encountering an abrupt transition to a charge of finite diameter or thickness, the detonation wave will undergo a decay in velocity to a steady state propagation velocity less than the CJ velocity, provided that the dimension (diameter or thickness) of the charge is greater than some critical value required for successful propagation. While there is a comprehensive literature devoted to determining the steady state detonation velocity in finite-sized charges and its dependence on the charge dimension, the decay or relaxation process has not been as extensively studied. The issue of the decay in velocity to steady state is frequently encountered in the design of experiments intended to measure the steady state propagation velocity in cylindrical charges (i.e., rate-stick tests), wherein the question of how long the wave must propagate in order to reach effectively constant velocity is encountered. The usual criterion used is that the wave must propagate on the order of five to ten diameters, although anecdotal accounts of charges that required significantly longer run distances to reach steady state can also be found. The origin of this criterion is unknown to the authors.

One of the earliest investigations of the propagation distance required to reach steady state is that of Cook et al.[1,2], who examined the radius of curvature of a point-initiated detonation (via a detonator) in cylinders of explosive. The radius of the wave initially grew linearly with propagation distance (following a geometrical spherical expansion) until, after encountering the edge of the explosive cylinder, the wave evolved to a constant curvature state. For the explosives studied by Cook et al., this process required a propagation distance of approximately five diameters to reach a steady state front curvature, which may be the origin of the criterion used in determining where steady state propagation begins in a rate-stick test. If these results are relevant to the relaxation of a planar-initiated detonation wave and how the decay process depends on the properties of the confinement, the diameter (or thickness) of the charge in comparison to the critical thickness, and on the wave structure of the detonation itself have not been quantified.

This paper uses computational simulations to investigate how an ideal CJ detonation decays to a steady state wave propagating in a finite diameter (or thickness) charge with yielding confinement. For simplicity, the ideal gas law is used for the equation of state and a pressure-dependent reaction rate (with pressure exponent $n = 2$ or $3$) is used to ensure a stable detonation wave structure. The results of the simulations are then compared to the predictions of the theory of Detonation Shock Dynamics (DSD) and a simple, heuristic model that attempts to describe the decay process in terms of rarefaction wave propagation.

**Computational Simulations**

The problem considered is shown schematically in Fig. 1. The detonation is initialized as a one-dimensional ZND-type wave propagating at the ideal CJ detonation velocity. The wave then propagates into a layer of finite thickness (in two-dimensional simulations) or a cylinder of finite diameter (in axisymmetric simulations) bounded by an inert medium. The explosive medium and confinement are both modeled as an ideal gas of equal molecular weight and ratio of specific heats $\gamma = 1.333$ (i.e., a single gas formulation). The impedance of the confinement was varied by varying the temperature and density of the confinement while maintaining the pressure as constant. In order to ensure the resulting detonation wave is stable (i.e., no cellular structure), a pressure-based reaction rate model is used:

$$\frac{\partial Z}{\partial t} = k(1-Z)\left(\frac{p}{p_{CJ}}\right)^n \quad (1)$$

where $Z$ is the reaction progress variable ($Z = 0$ for unreacted and $Z = 1$ for fully reacted) and the values of $n = 2$ and $3$ are used.

The unsteady, two-dimensional Euler equations were solved the coupled with a pressure-dependent reaction rate by using a second-order accurate algorithm. In order to isolate the stiff source term of chemical reaction, a second-order accurate Strang operator splitting method[3] was employed. The AUSM+ scheme[4] was used to deal with the inviscid flux as a sum of the convective and pressure terms, recognizing the convection and acoustic waves as two physically distinct processes. A third-order TVD Runge-Kutta method[4] was used for the temporal discretization. The boundary condition along the $x$-axis was a mirror boundary condition (axis of symmetry), so that only the upper half of the layer was simulated in the case of a two-dimensional slab. The upper boundary of the computational domain (above the inert layer) was a supersonic outflow condition to ensure that no reflected waves return into the computational domain. A geometric source term was included in axisymmetric calculations to simulate detonation propagation in a cylindrical charge of explosive surrounded by an annular layer of inert confinement.

The simulations were conducted in the laboratory reference frame, with the computational domain periodically shifted to prevent the detonation from encountering the end of the domain. The downstream region of the domain that was truncated was always located beyond the location of the limiting characteristic following the detonation to ensure that there was no influence on the wave dynamics by this process. A thorough grid resolution study was conducted and determined that the global dynamics of the wave are not influenced as the grid resolution was increased beyond 5 computational cells per half reaction zone length of the ideal CJ detonation; the details of this resolution study are given in Ref. 5.

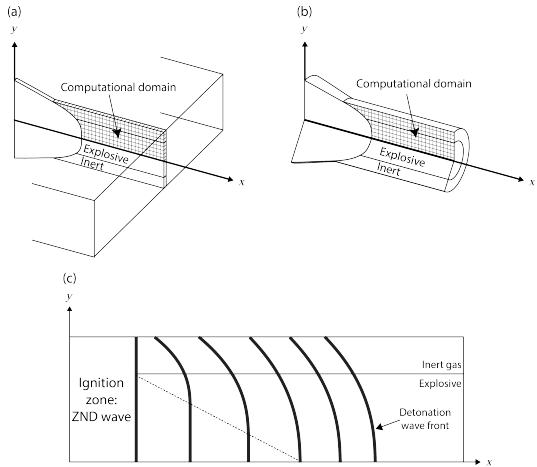

Fig. 1 Schematic representation of the problem to be studied showing the computational grid for (a) two-dimensional slab and (b) axisymmetric geometries, and (c) regions of simulation.

Unless otherwise noted, the simulations reported here were performed at a resolution of 5 cells per half reaction zone length.

The results were post-processed to extract the dynamics of the detonation front. For each output data file, the location where the density along the centerline of the charge (the lower boundary of the simulations) increased to twice the initial density of the explosive was considered as the position of the leading shock front at that time. The data of leading shock position *vs*. time was then converted to detonation velocity *vs*. time via a central differencing scheme.

The results of the simulations are shown in Fig. 2. Figure 2(a) shows the effect of varying the charge thickness in a two dimensional simulation of different thicknesses with pressure exponent $n = 2$ and the impedance of the confinement equal to that of the explosive. The time since the detonation front first encounters the finite thickness or diameter charge has been normalized by a characteristic time, $t_c$, for a fluid particle passing through the reaction zone, which is defined as the half reaction zone thickness of the ideal CJ detonation ($L_{1/2}$) divided by the von Neumann fluid velocity with respect to the leading wave front, and the detonation velocity normalized by the ideal CJ detonation velocity. The wave is seen to continue to propagate at constant velocity along the centerline for some distance prior to

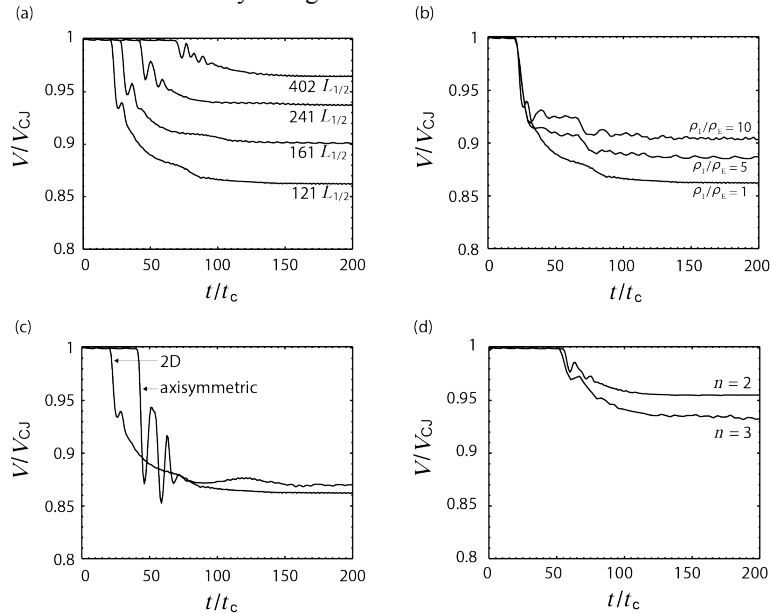

Fig. 2 Detonation velocity (normalized by ideal CJ velocity) as a function of time (normalized by characteristic time), showing the effect of varying (a) charge thickness, (b) impedance of the confinement, (c) charge geometry, and (d) pressure exponent on the decay of detonation wave from its initial CJ velocity to the steady state terminal velocity.

beginning the decay process. The arrival of this initial disturbance along the centerline is treated in detail later in the paper. The larger thickness charges exhibit less of a velocity deficit because of the comparatively smaller losses in momentum due to lateral expansion of the products into the yielding confinement. The final steady state velocity for the case in Fig. 2 is shown to correspond well to a simple curvature based model originating with Eyring et al.[5] and using a normal detonation velocity-curvature ($D_n$-$\kappa$) relation obtained by integrating the differential equation of Wood and Kirkwood.[6] This model can be shown to be equivalent to the predictions of DSD if the same $D_n$-$\kappa$ relation is used; additional details are given in Ref. 7

Figure 2(b) shows the effect of varying the impedance of the confinement by changing the density and (inversely) the temperature of the inert gas at constant pressure for a two dimensional slab of thickness 121 $L_{1/2}$. The higher impedance (denser) confinement exhibits less of a velocity deficit due to the greater confinement of the products provided by the higher impedance inert layer. In the case of high impedance confinement, an even simpler one-dimensional stream tube model that uses Newtonian hypersonic impact theory to model the interaction with the confinement has proven to be very successful in predicting the steady state velocity.[7]

Figure 2(c) shows an example of the results for axisymmetric simulations (cylindrical charges) for the same conditions as in Fig. 2(a). The simulation is presented with the diameter of the axisymmetric charge paired with charge of thickness (121 $L_{1/2}$) equal to half that diameter, since cylindrical charges of diameter equal to twice the thickness of a slab charge are expected to reach the same steady state velocity. This 2:1 scaling between diameter and thickness was verified in the study of Li et al.[5] A steeper initial deceleration of the wave can be seen in the axisymmetric case, followed by a more oscillatory behavior in comparison to the analogous planar case. Figure 2(d) shows the effect of varying the pressure exponent from $n$ = 2 to 3 in a two dimensional simulation of the same normalized thickness (321 $L_{1/2}$) and the impedance of the confinement equal to that of the explosive.

The results shown here are only a subset of all simulations performed for purposes of clarity. Attempts by the authors to collapse all of these results onto a single curve (i.e., assuming the decay process is self-similar when scale by charge or reaction zone thickness) were unsuccessful. This non-self-similarity property can be attributed to the existence of multiple length scales in the problem (the detonation reaction zone thickness and the charge dimension).

**Detonation Shock Dynamics**

In order to treat problems involving transient propagation of detonation fronts without having to perform computationally intensive direct simulations, the theory of Detonation Shock Dynamics (DSD) has been developed over the last three decades.[8,9] DSD treats detonation dynamics via a front evolution equation that postulates a known relationship between propagation of the detonation normal to the front and the local curvature of the front ($D_n$-$\kappa$ relation). The location of the front is determined by the contour of a parabolic (diffusion-like) partial differential equation subject to boundary conditions. The result is similar to a classical Huygens construction for front propagation, but with the propagation velocity properly reflecting the slower velocity of a curved front. This method was intended to solve problems similar to those considered in this paper, but is subject to the limitation that the reaction zone must be thin in comparison to the radius of curvature of the front and the other length scales involved in the problem of interest. One of the first applications of DSD theory was to solve the same problem considered in this paper, namely, the decay of a detonation initialized as a CJ detonation to its steady state condition.[10]

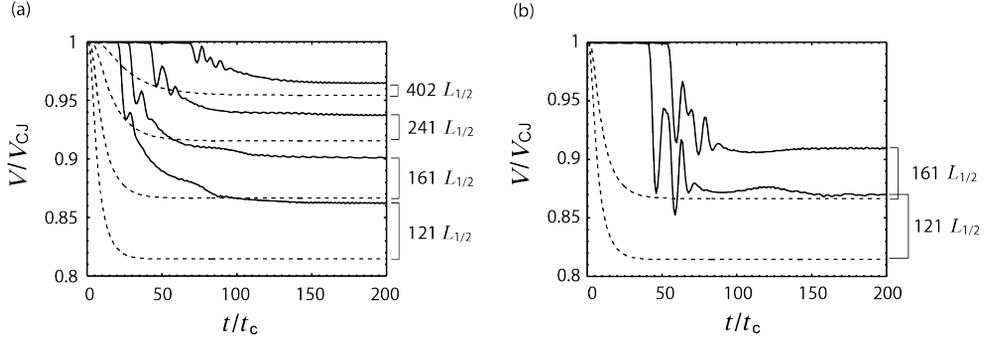

Fig. 3 Detonation velocity (normalized by ideal CJ detonation velocity as a function of time (normalized by characteristic time) compared to the predictions of DSD theory for equal impedance confinement and $n = 2$ in (a) two dimensional slab and (b) axisymmetric geometries. The solid lines are the results of simulation. The dashed lines are the predictions of DSD theory.

With the $D_n$-$\kappa$ relation known, the evolution of the leading shock front in a two-dimensional planar space, $x_s(t,y)$ throughout the decay process can be obtained by solving the following partial differential equation,

$$\frac{\partial x_s}{\partial t} = D_N(\kappa)\sqrt{1+\left(\frac{\partial x_s}{\partial y}\right)^2} - D_{CJ} \quad (2)$$

where, by definition, $\kappa$ is

$$\kappa = \frac{\partial^2 x_s}{\partial y^2} \bigg/ \left[1+\left(\frac{\partial x_s}{\partial y}\right)^2\right]^{\frac{3}{2}} \quad (3)$$

The boundary condition required to solve Eq. (2), i.e., the angle of the leading shock front at the boundaries of the charge, can be determined by shock polar analysis. Additional details of this technique can be found in Ref. 7. With the applied boundary condition and $D_n$-$\kappa$ relation fitted by a fourth-order polynomial, Eqs. (2) and (3) were solved numerically using the method of lines.

In Fig. 3, the predictions of DSD theory of the detonation velocity along the centerline are compared to the results of numerical simulations for both two-dimensional and axisymmetric geometries at various thicknesses or diameters with the impedance of the confinement equal to that of the explosive. The results shown in Fig. 3 indicate poor agreement between the predictions of DSD theory and the results of the numerical simulations reported in the previous section. This discrepancy is not due to the $D_n$-$\kappa$ relation used in implementing the DSD theory. The authors have previously shown that the steady propagation velocity and front curvature measured from the simulations match the predicted $D_n$-$\kappa$ relation obtained by integrating the Wood and Kirkwood model along the central streamline.[7] Rather, the major discrepancy of the DSD model can be attributed to its inability to capture the finite signaling time required for the initial rarefaction to reach the charge centerline. This hyperbolic (wave-like) phenomenon cannot be reproduced in the parabolic (diffusion-like) DSD evolution equation (Eq. (2) above) that results in an infinite propagation velocity.

To address this deficiency, an improved DSD theory was developed that postulated a relation between the detonation velocity, the rate of change of detonation velocity, and the curvature ($D_n$-$\dot{D}_n$-$\kappa$).[11,12] The governing equation in this improved model exhibits both hyperbolic and parabolic behavior, so it can successfully reproduce the finite propagation velocity of the leading rarefaction that signals the existence of the yielding confinement.[11,12] This improved model, however, requires solution of the governing PDE via numerical finite difference techniques, a task which can be as challenging as direct numerical simulation of the Euler equations. In addition, there is not yet a theory able to generate the required $D_n$-$\dot{D}_n$-$\kappa$ relation, and when implemented, this relation has been fit to direct numerical simulations.[11,12] Furthermore, as shown in Fig. 3(b), solving the DSD governing equation for in a two-dimensional planar space fails to capture the steeper initial deceleration observed in the results of axisymmetric simulations. In order to model the evolution of detonation wave fronts in a cylindrical charge, the DSD equation needs to be solved in three-dimensional space or using an empirical $D_n$-$\dot{D}_n$-$\kappa$ relation for the axisymmetric geometry. Thus, this approach is not conducive to simple estimates of the relaxation distance required to reach steady state propagation, and as such, will not be explored further here.

**Heuristic Model**

In order to provide a model that can be used for quick, engineering calculations of the distance required to reach steady state, the phenomena observed in the simulations will be described by models that reflect the gas dynamics occurring in the relaxation process.

Leading Rarefaction

The detonation wave will propagate along the centerline of the charge at the ideal CJ velocity until the information of yielding confinement is communicated to the wave front at that location. A theoretical picture which describes this process was originally developed by Skews[13] in the study of diffracting shock waves and applied to diffracting detonation waves by Schultz[14]. As illustrated in Fig. 4, the disturbance of yielding confinement is transmitted to the wave front at the centerline of the charge at the local speed of sound, $c$, while being advected at velocity $u$ in a laboratory-fixed frame of reference. Hence, the transverse signaling speed, $v$, is determined using the velocity triangle in Fig. 4 as

$$v = \sqrt{c^2 - (D_{CJ} - u)^2} \quad (4)$$

The leading rarefaction propagates into the detonation flow at the maximum value of $v$ within the reaction zone of the ZND detonation structure. By solving the governing equations for ZND detonation, $c$ and $u$ can be obtained as functions of space within the reaction zone. Then, the spatial profile of $v$ can be calculated based on the solutions for $c$ and $u$, and a maximum can be identified within this profile.

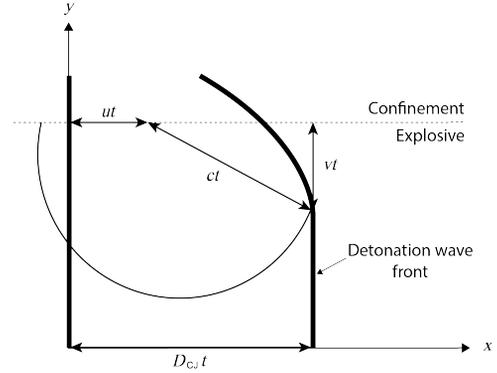

Fig. 4 Illustration of the leading disturbance from yielding confinement transversely propagating into the explosive.

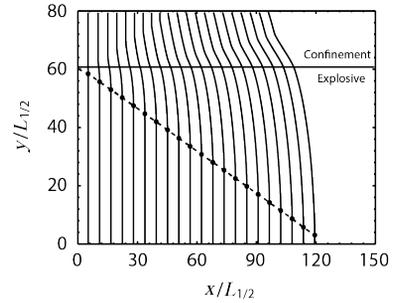

Fig. 5 Evolution of detonation wave front (with spatial coordinates normalized by $L_{1/2}$) for a two dimensional slab of thickness 121 $L_{1/2}$ with equal impedance confinement and $n$ = 2. The solid curves are the extracted wave fronts at different times. The dashed line is the trajectory of the leading disturbance before reaching the centerline of the charge.

Despite the accepted nature of this picture, we are not aware that this model of the leading rarefaction has been rigorously verified by numerical simulations. By applying the same wave front tracking method to the entire computational domain, the complete profile of the leading shock front was extracted for each output data file. As shown in Fig. 5, a point separating the disturbed and undisturbed regions was identified on each profile. By calculating the slope of the straight line connecting these separating points, the numerical results of the leading rarefaction speed were obtained. As shown in Fig. 6, as the grid resolution was increased to 15 computational cells per half reaction zone length, the numerical result of the leading rarefaction speed converged to the value predicted by Eq. (4) evaluated using the ZND structure of the ideal CJ detonation. In order to account for the uncertainty in finding the location of the leading disturbance, error bars defined by the computational grid size divided by the time interval between two consecutive output data files are plotted with the numerical data points in Fig. 6.

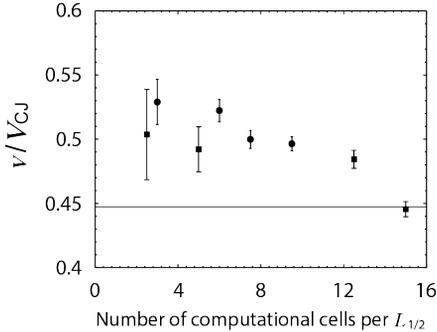

Fig. 6 Numerical results of the leading rarefaction speed (normalized by the ideal CJ velocity) for pressure exponent $n = 2$ (square) and $n = 3$ (circle) as a function of number of computational cells per $L_{1/2}$. Solid line is maximum signal speed of ideal ZND wave.

Rarefaction Gradient and Wave Deceleration

Beyond the leading rarefaction, the detonation deceleration becomes a complex process of unsteady and multidimensional gas dynamics as the wave accommodates itself to the yielding confinement. It would be unrealistic to expect a simple, analytic model can capture the full details of this process, however, a phenomenological model that includes the correct physics may have some success in describing the relaxation rate. This model is based on the following assumptions:
1. The initial wave deceleration along the center following the arrival of the leading rarefaction is determined by the gradient of the rarefaction.
2. The rarefaction gradient can be approximated by computing a one-dimensional shock tube problem in the direction perpendicular to the direction of detonation propagation, with the "driver" conditions being taken as the von Neumann (post-shock) state of the ideal CJ detonation and the "driven" state being the confinement.
3. The detonation continues to decay at the rate dictated by the initial rarefaction gradient until it reaches the steady state velocity. The steady state propagation velocity is assumed to be known. A number of models, with varying degree of sophistication, are available to predict the steady state velocity for a given explosive and confinement. For the system studied here, these models are presented and compared in Ref. 7.

Under assumption (2), the initial rarefaction gradient is independent of the confinement material. In a shock-tube problem, the magnitude of the initial rarefaction gradient is independent of the driven medium that the shock propagates into. Only the duration of the rarefaction (i.e., how long the rarefaction continues) changes with varying conditions in the driven state. This assumption appears to be supported by the results shown in Fig. 3, which shows that the initial deceleration of the wave is the same as the density of the confinement is varied by a factor of 10.

Once the leading rarefaction reaches the centerline of the charge, the rarefaction waves emanating from the yielding confinement on both sides of the charge interact with each other and form a non-simple wave region. This non-simple rarefaction wave reduces the static pressure behind the leading shock wave and thus causes the leading shock to decelerate. The flow condition within this non-simple rarefaction wave at the centerline of the charge only depends on the initial rarefaction gradient and can be analytically determined by the method of characteristics.[15] By using Riemann's solution as implemented by Rudinger (see details in Ref. 15), the post-shock pressure influenced by the interacting rarefaction waves at the centerline of the charge can be obtained as a function of time.

For the case of a cylindrical charge, the inward propagating rarefaction results in an initially greater rate of deceleration due to the geometric factor associated with the axisymmetric geometry. A general analytic solution for the inward propagating rarefaction was provided by Greenspan and Butler in the study of the isentropic expansion of a cylindrical and spherical gas cloud into a vacuum[16], and later corrected and extended by Rasmussen and Frair.[17] By implementing the solution from Rasmussen and Frair (see details in Ref. 17) for the case of $\gamma = 1.333$, the normalized speed of sound, $c/c_o$, can be obtained as a function of normalized position, $r/r_o$ at the moment when the leading rarefaction reaches the center of the charge for both planar and axisymmetric geometries, i.e., when the normalized time $c_o t/r_o = 1$. The initial rarefaction gradient for planar and cylindrical geometries can be determined by evaluating the derivative of $c/c_o$ with respect to $r/r_o$ at $r/r_o = 0$ and $c_o t/r_o = 1$ as follows,

$$\frac{d(c/c_o)}{d(r/r_o)}\bigg|_{plan.} = 0.1429 \quad (5)$$

$$\frac{d(c/c_o)}{d(r/r_o)}\bigg|_{cyl.} = 0.2055 \quad (6)$$

Since the rarefaction is an isentropic process,

$$\frac{P}{P_o} = \left(\frac{T}{T_o}\right)^{\frac{\gamma}{\gamma-1}} = \left(\frac{c}{c_o}\right)^{\frac{2\gamma}{\gamma-1}} \quad (7)$$

For $\gamma = 1.333$, a geometric factor, which can be applied to Rudinger's solution for the pressure at the centerline of a two-dimensional planar charge during the interaction of rarefaction waves to approximate the solution in an axisymmetric case, was determined as follows,

$$\alpha_{cyl.} = \left(\frac{0.1429}{0.2055}\right)^8 = 0.0547 \quad (8)$$

Up to this point, only the unreactive gasdynamic aspect of the problem has been considered. In fact, the exothermic reaction acts like a piston to drive the leading shock and tends to maintain the post-shock pressure at the von Neumann condition. Hence, it is essential to consider the interaction between the rarefaction wave reducing the post-shock pressure and the exothermic reaction maintaining the post-shock pressure. In this heuristic model, the effect of the reaction zone is simplified by introducing a parameter of reaction influence, β, which satisfies the following relation:

$$P_{ps}(t) = \beta P_{ps}^*(t) + (1-\beta)P_{vN} \quad (9)$$

where the subscript "ps" denotes the post-shock condition, the superscript "*" denotes the solution for unreactive gas dynamics, and $P_{vN}$ is the von Neumann pressure in the ideal ZND detonation. According to Eq. (5), the actual value of $P_{ps}$ is between the unreactive post-shock pressure and $P_{vN}$ as β varies from 0 to 1. As mentioned in assumption (3), the steady state to which the detonation decays can be successfully predicted by a number of independent models. Hence, β can be determined by applying the steady states condition as follows,

$$\beta = \frac{P_{vN} - P_{ss}}{P_{vN} - P_{ps}^*(t_{end})} \quad (10)$$

where $P_{ss}$ is the post-shock pressure for the steady state detonation.

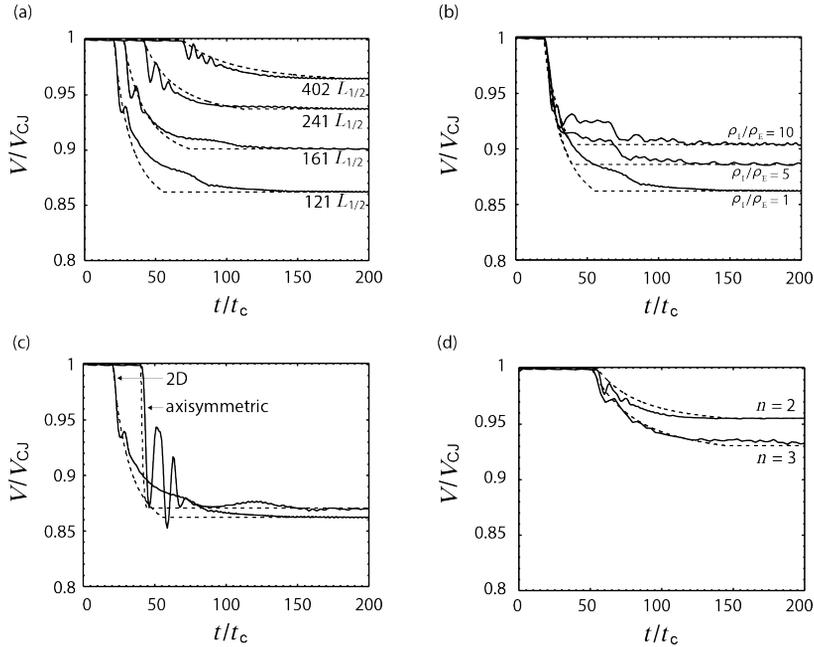

Fig. 7 Detonation velocity (normalized by ideal CJ detonation velocity) as a function of time (normalized by characteristic time) compared to the predictions of the heuristic model, showing the effect of varying (a) charge thickness, (b) impedance of the confinement, (c) charge geometry, and (d) pressure exponent on the decay of a detonation wave from its initial CJ velocity to the steady state terminal velocity. The dashed lines are the predictions of the heuristic model.

The decays of detonation velocity predicted by this heuristic model are compared to the results of simulations in Fig. 7. The model predictions of detonation velocity shown in Fig. 7 are functions of time, $V(t)$, which can be converted to functions of space, $V(x)$, via the following relation,

$$x = \int_0^t V(\tau)d\tau \quad (11)$$

where τ is a dummy variable representing $t$.

Figure 7(a) compares the model predictions to the numerical results shown in Fig. 2(a). It is important to note that the predictions agree well with numerical results for large thicknesses of the charge, however, the model deviates from the numerical results as thickness decreases. The explanation is that, for smaller thicknesses, the decay process is not only influenced by the initial rarefaction waves from yielding confinement, but also by those reflected back into the charge from the material interface between the explosive and the confinement media.

Figure 7(b) compares the model prediction for various impedances of the confinement with the results of simulations shown in Fig. 2(b). The model well captures the feature that the decay rate of detonation velocity is independent of the impedance of the confinement. The only significant influence of the confinement is on the steady state velocity. Figure 7(c) compares the model prediction for two-dimensional slab and axisymmetric geometries with the results of simulations shown in Fig. 2(c). By considering a larger gradient of the initial rarefaction waves in cylindrical charges, the model fairly well predicts a steeper deceleration in axisymmetric geometry than that in two-dimensional slab. Figure 7(d) compares the model prediction for different pressure exponents, $n = 2$ and 3, with the results of simulations shown in Fig. 2(d). Although the initial rarefaction gradient is expected to be the same for $n = 2$ and 3, the model predicts a slightly steeper decay for $n = 3$ which has a shorter half reaction zone length.

**Conclusions**

The results of the present investigation reiterate that problems in unsteady detonation dynamics are inherently problems of wave propagation, and correct description of the detonation relaxation process must include the signaling via rarefaction waves generated by the interaction with the confinement. The original version of DSD theory, using a parabolic front evolution equation, fails to include this feature, which has motivated more sophisticated models that include the rate of detonation deceleration. However, until a theory that can predict the relation between detonation velocity, the rate of change of velocity, and the front curvature ($D_n$-$\dot{D}_n$-κ) is developed, this model requires calibration via numerically intensive, direct numerical simulation.

A simpler, heuristic model proposed here, which treats the interaction with the confinement via a one-dimensional shock-tube like problem occurring orthogonal to the direction of propagation, appears to have some promise for quick calculation of the detonation decay rate to steady state. The model would require further verification using numerical simulations with equations of state more representative of actual explosives as well as comparison to experimental results in order for a comprehensive assessment of its usefulness to be made.

**Acknowledgements**

This work was supported by the National Natural Science Foundation of China (No. 51206150), National Key Laboratory for Shock Wave and Detonation Physics Research Foundation (No. 9140C6704010704) and the State Scholarship from China Scholarship Council.